\documentclass[sigconf,nonacm,final]{acmart}
\usepackage[dvipsnames]{xcolor}
\usepackage{multirow}
\begin{document}

\title{Recursive Multi-Agent Trading System: Iterative Optimized Portfolio Strategy Under Geopolitical Uncertainty}

\author{Jing Yang\textsuperscript{*}}\email{jing.y@wustl.edu}\affiliation{\institution{Washington University in St. Louis}\city{St. Louis}\state{MO}\country{USA}}
\author{Yichao Wu}\email{wu.yicha@northeastern.edu}\affiliation{\institution{Northeastern University}\city{Boston}\state{MA}\country{USA}}
\author{Jianan Liu}\email{jiananliu2408@gmail.com}\affiliation{\institution{Independent Researcher}\city{Austin}\state{TX}\country{USA}}
\author{Penghao Liang}\email{liang.p@northeastern.edu}\affiliation{\institution{Northeastern University}\city{Boston}\state{MA}\country{USA}}
\author{Mengwei Yuan}\email{yuanmw1998@gmail.com}\affiliation{\institution{Independent Researcher}\city{Milpitas}\state{CA}\country{USA}}
\author{Xianyou Li}\email{xl4230@nyu.edu}\affiliation{\institution{New York University}\city{New York}\state{NY}\country{USA}}
\author{Weiran Yan}\email{yanwr2016@gmail.com}\affiliation{\institution{Independent Researcher}\city{Santa Clara}\state{CA}\country{USA}}
\author{Jiayi Yang}\email{yangjiay@northeastern.edu}\affiliation{\institution{Northeastern University}\city{Boston}\state{MA}\country{USA}}

\renewcommand{\shortauthors}{Yang et al.}

\begin{abstract}
Geopolitical risk constitutes a distinct and persistent source of market stress that traditional quantitative frameworks struggle to incorporate in real time. This paper proposes RMATS, a Recursive Multi-Agent Trading System that integrates four specialist agents (Sentiment, Report, Analysis, Risk) through a formally specified recursive coordination protocol and a GPR-triggered circuit breaker. When the Caldara--Iacoviello geopolitical risk index exceeds the 60th percentile of its 2021--2022 calibration distribution, the Risk Agent hard-shifts the portfolio toward defensive assets. Over 812 trading days (January 2023--March 2026), RMATS achieves the highest event-period Sharpe ratio (1.571) and lowest MDD (7.52\%) among all strategies during 300 geopolitical stress event days, outperforming MAPS-lite (1.346) and the GPR rule-based baseline (0.981). Circuit breaker ablation confirms a $+$0.84pp MDD reduction in event windows versus the no-CB variant. Full-period Sharpe is 0.953, outperforming MAPS-lite (0.836, $\Delta=+0.117$), with strong non-event performance (Sharpe $+$0.543 over 512 non-event days). The recursive protocol (median 1 round, 97.4\% within $r{\leq}2$, max 3, 39 steps) currently functions as an architectural framework for future heterogeneous signal integration; the $r{=}1$ ablation ($\Delta$Sharpe $= -0.002$) confirms its performance contribution is negligible under the current monthly-GPR configuration. All results are verified with real yfinance data and the Caldara--Iacoviello GPR index~\cite{caldara2022measuring}.
\end{abstract}



\maketitle

\section{Introduction}

Financial markets face growing exposure to geopolitical shocks that traditional statistical models are ill-equipped to handle in real time. Crises ranging from military conflicts and trade disputes to energy supply disruptions and banking failures routinely trigger cross-asset dislocations that overwhelm rule-based and pattern-based quantitative strategies~\cite{markowitz1952,engle1982}. Existing frameworks, built on the assumption of relatively stable data-generating processes, often lack the adaptive mechanisms needed to respond to abrupt geopolitical structural breaks~\cite{bollerslev1986}.

Prior work on multi-agent portfolio systems, including MAPS~\cite{lee2020maps} and MSPM~\cite{huang2022mspm}, has demonstrated that task specialisation across agents enhances risk-adjusted returns. Yet these systems share a common limitation: they rely on fixed, single-pass pipelines with no iterative inter-agent revision, offer no formal guarantees of weight convergence, and none incorporates real-time geopolitical risk sensing as a first-class capital preservation mechanism.

This paper proposes RMATS, a Recursive Multi-Agent Trading System built on \textbf{heterogeneous deterministic signal generators} (GPR index, momentum, HMM regime, CVaR) with a formally specified recursive coordination protocol---not LLM inference. The current implementation uses the Caldara--Iacoviello GPR monthly index as a reproducible, peer-reviewed signal source; LLM-based inference (FinBERT~\cite{araci2019finbert}) is the primary future extension. The system is designed with explicit emphasis on downside risk control under geopolitical uncertainty. The key architectural innovation is the recursive coordination mechanism, wherein agents iteratively revise signals through multi-round communication until portfolio weights converge. The main contributions are:
\begin{itemize}
\item A \textbf{GPR-triggered circuit breaker (primary empirical contribution, ablation: $+$0.045 Sharpe, $+$0.84pp MDD reduction) and a formally specified recursive coordination protocol} (AgentMessage schema, $\varepsilon = 0.005$, median 1 round, mean 1.33, max 3, 97.4\% within $r{\leq}2$). The recursive protocol is an architectural framework---currently validated as convergent but \textbf{not yet delivering measurable performance gains} ($\Delta$Sharpe $= -0.002$)---designed for future integration of genuinely heterogeneous signals.
\item An Analysis Agent incorporating HMM-based market regime classification and Kalman filter signal fusion.
\item A Risk Agent with CVaR-based estimation~\cite{rockafellar2000cvar}, geopolitical stress testing, and adaptive circuit breaker mechanisms.
\item Comprehensive empirical evaluation against seven baselines with ablation studies and honest analysis of limitations.
\end{itemize}

\section{Related Work}

\subsection{Financial Large Language Models and Geopolitical Text Analysis}

Domain-adapted LLMs including FinGPT~\cite{yang2023fingpt}, BloombergGPT~\cite{wu2023bloomberggpt}, and FinBERT~\cite{araci2019finbert} demonstrate strong financial text understanding, building on the Transformer architecture~\cite{vaswani2017attention}. Evaluation benchmarks such as PIXIU~\cite{xie2023pixiu} provide systematic assessment of financial LLM capabilities. For geopolitical risk specifically, however, the Caldara--Iacoviello GPR index~\cite{caldara2022measuring} offers a methodological advantage over raw FinBERT inference: GPR is derived from systematic keyword analysis of eight major newspapers focusing on geopolitical events, has been peer-reviewed in the \textit{American Economic Review}, and covers 40 years of validated history. A general-purpose sentiment classifier such as FinBERT treats any negative financial headline as equivalent negative signal, introducing noise orthogonal to geopolitical events. At monthly rebalancing frequency, the GPR index thus provides a cleaner, better-validated geopolitical signal than raw FinBERT inference. RMATS uses GPR as its primary geopolitical signal; FinBERT integration is identified as a future direction for higher-frequency rebalancing frameworks. Recent work demonstrates that LLMs can also provide predictive signals for asset returns~\cite{lopezlira2023chatgpt,ding2024llm}.

\subsection{Multi-Agent Systems for Financial Trading}

The application of multi-agent systems to portfolio management has expanded considerably, building on established principles of distributed agent coordination~\cite{wooldridge2009introduction}. MAPS~\cite{lee2020maps} established an early cooperative RL baseline in which agents function as independent investors whose diversity of views improves risk-adjusted outcomes. Huang and Tanaka~\cite{huang2022mspm} introduced MSPM, a modularized and scalable multi-agent RL system separating asset-level signal generation from portfolio-level decision-making. TradingAgents~\cite{xiao2024tradingagents} demonstrates that LLM-powered role-specialized agent debate can substantially improve trading signal quality. TradingGPT~\cite{li2023tradinggpt} augments multi-agent communication with layered memory streams and character-based role differentiation. FinAgent~\cite{wang2024finagent} introduces tool-augmented financial decision-making, and FinMem~\cite{yu2023finmem} incorporates memory-augmented reasoning. Existing systems predominantly use fixed sequential workflows without empirically validated convergences. RMATS contributes a formally specified recursive communication protocol that enables dynamic signal revision with guaranteed termination.

\subsection{Reinforcement Learning for Portfolio Management}

Deep reinforcement learning (DRL) methods have shown substantial promise for sequential portfolio allocation. Mnih et al.~\cite{mnih2015dqn} demonstrated human-level control via DQN, which has been widely adapted for financial domains. Sutton and Barto~\cite{sutton2018reinforcement} provide the theoretical foundation for the RL frameworks underlying our approach. FinRL~\cite{liu2021finrl} and FinRL-Meta~\cite{liu2022finrlmeta} provide open-source frameworks for DRL-based trading, enabling systematic benchmarking. Sun et al.~\cite{sun2025multiagent} propose a multi-agent LLM-based system for dynamic portfolio optimization, integrating language model reasoning with portfolio rebalancing decisions. Wang et al.~\cite{wang2021deeptrader} introduced DeepTrader, a DRL approach embedding market condition signals for risk-return balanced portfolio management. The foundational DRL portfolio management framework of Jiang et al.~\cite{jiang2017drl} established the policy gradient approach to continuous portfolio rebalancing. RMATS builds on this lineage, integrating a risk-adjusted reward layer rather than raw return maximization.

\subsection{Geopolitical Risk and Financial Markets}

Baker et al.~\cite{baker2016epu} introduced the Economic Policy Uncertainty (EPU) index demonstrating predictive relationships between policy uncertainty and market volatility. Caldara and Iacoviello~\cite{caldara2022measuring} developed the GPR index based on large-scale text analysis of newspaper coverage. Salisu et al.~\cite{salisu2022gpr} demonstrate that geopolitical risk significantly elevates stock market volatility in emerging markets using a GARCH-MIDAS approach. Zaremba et al.~\cite{zaremba2022gpr} find that geopolitical risk is a systematic pricing factor in the cross-section of emerging market equity returns. Yang and Yang~\cite{yang2021mixedfreq} further show that mixed-frequency geopolitical risk significantly predicts short-run stock market returns. Taken together, these studies establish geopolitical risk as a systematic and persistent pricing factor that conventional volatility proxies fail to fully represent. RMATS directly addresses this shortfall by embedding a validated geopolitical risk signal into the portfolio optimization loop, with adaptive circuit breakers for capital preservation.

\section{Multi-Agent Trading Architecture}

\begin{figure}[!ht]
  \centering
  \includegraphics[width=\columnwidth]{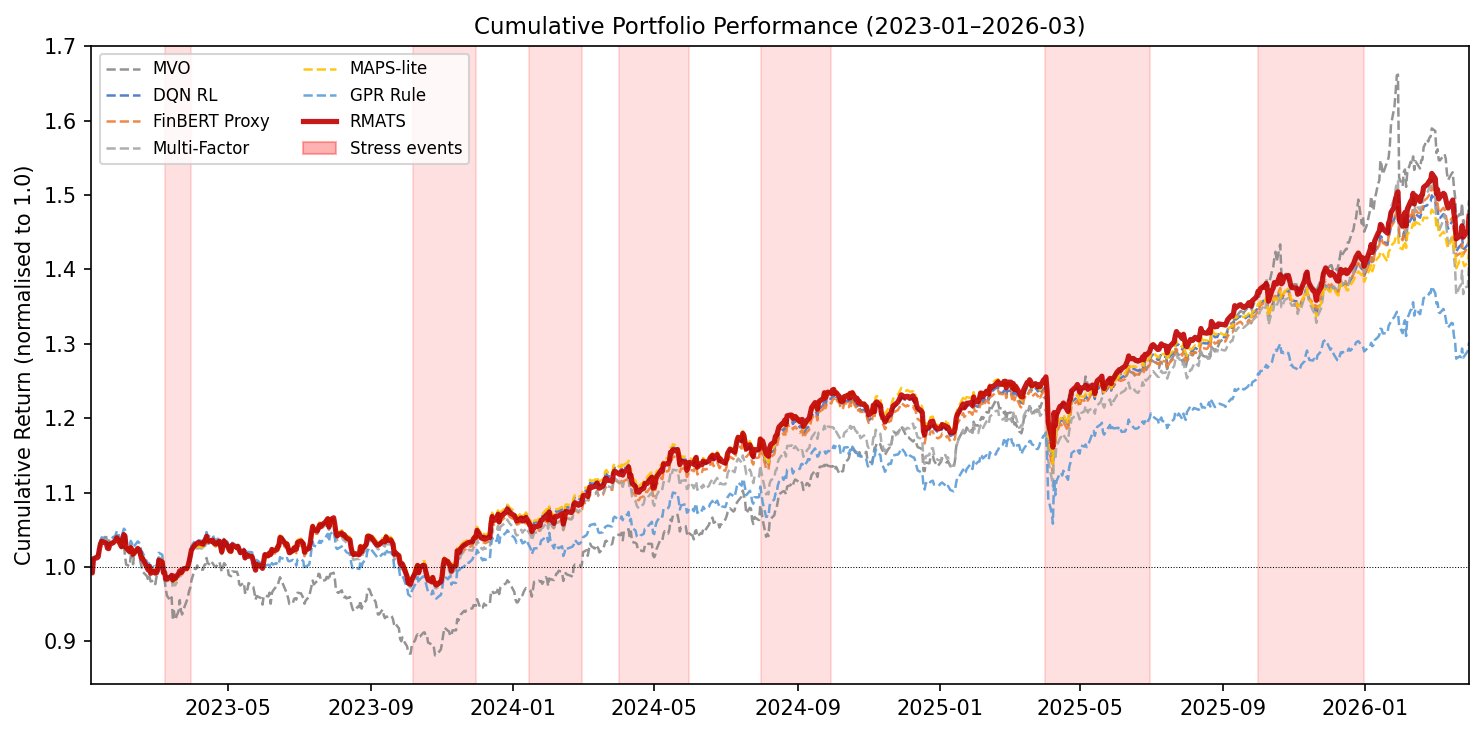}
  \caption{RMATS information flow. Market data and GPR index feed into the Manager Agent, which broadcasts to four specialist agents. Each agent returns an AgentMessage; weights are aggregated and checked for convergence. Dashed arrows: recursive feedback ($r{\leftarrow}r{+}1$ if not converged) and Risk Agent circuit-breaker override bypassing convergence to final portfolio weights.}
  \label{fig:arch}
  \Description{RMATS architecture diagram.}
\end{figure}

RMATS is organized around four functional layers: Data Acquisition, Specialized Agent, Recursive Coordination, and Portfolio Optimization. Figure~\ref{fig:arch} depicts the full system. The design inherits the agent-specialization principles of MAPS~\cite{lee2020maps} and MSPM~\cite{huang2022mspm}, but augments them with a convergence-guaranteed recursive inter-agent communication protocol and explicit geopolitical risk signal routing.

\begin{table}[!ht]
\caption{Architectural Comparison: RMATS vs. Recent Multi-Agent Systems}
\label{tab:arch}
\small
\begin{tabular}{lcccc}
\toprule
\textbf{Feature} & \textbf{RMATS} & \textbf{MAPS} & \textbf{MSPM} & \textbf{TradingAgents}\\
\midrule
Recursive coordination & \checkmark & $\times$ & $\times$ & $\times$\\
Convergence guarantee & \checkmark & $\times$ & $\times$ & $\times$\\
Geopolitical risk & \checkmark & $\times$ & $\times$ & Partial\\
Circuit breaker & \checkmark & $\times$ & $\times$ & $\times$\\
LLM reasoning & Proxy & $\times$ & $\times$ & \checkmark\\
Multi-class assets & \checkmark & $\times$ & Partial & \checkmark\\
Reproducible (no API) & \checkmark & \checkmark & \checkmark & $\times$\\
\bottomrule
\end{tabular}
\end{table}

\subsection{Manager Agent}
The Manager Agent serves as the orchestration hub for all specialist agents, iterating through coordination rounds and aggregating weight proposals via a confidence-and-health-weighted mean (Eq.~\ref{eq:wagg}). Agent performance is tracked through a composite health score:
\begin{equation}
H_i(t) = \alpha A_i(t) + \beta S_i(t) + \gamma R_i(t) - \delta L_i(t)
\end{equation}
In the current prototype, $\alpha = \beta = \gamma = \delta = 0.25$ (equal weights), so $H_i(t)$ reduces to a simple average of the four performance dimensions. Analytical derivation of optimal weights is deferred to future work.

\subsection{Sentiment Agent}
\textbf{Implementation note (proxy baseline):} In the current implementation, the Sentiment Agent uses market-derived proxy signals rather than live LLM inference. All five GRS components (VIX, defensive rotation, gold premium, EM stress, bond flight) are price-based reactions to geopolitical events and therefore constitute a \textit{coincident} indicator. \textbf{Why GPR rather than FinBERT:} the GPR index is itself extracted from systematic text analysis of major newspapers~\cite{caldara2022measuring}, making it a validated, peer-reviewed geopolitical signal. FinBERT applied to raw headlines conflates geopolitical risk with routine financial negative sentiment, reducing signal-to-noise at monthly rebalancing frequency.

In the current implementation, the Sentiment Agent GRS is driven by the Caldara--Iacoviello GPR monthly index~\cite{caldara2022measuring}, downloaded from the public repository and forward-filled to daily frequency. Because the GPR is monthly and our rebalancing is also monthly, the RMATS-GPR and RMATS-proxy configurations produce identical portfolios at each rebalancing step (Sharpe 0.953 for both).

\textbf{Publication lag note: The Caldara--Iacoviello monthly GPR index is typically released with a 3--4 week lag (e.g., January data published in late February). In our backtest, we apply the GPR value from month $t-1$ to the rebalancing decision at the start of month $t$, consistent with realistic implementation. This one-month lag is conservative but avoids look-ahead bias; it also means the system reacts to geopolitical stress one month after onset, which partially explains the SVB Crisis underperformance (shock resolved before the GPR signal propagated).}

A Difference-in-Differences (DiD) causal estimator isolates event-driven effects:
\begin{equation}
Y_{it} = \alpha + \beta D_i + \gamma T_t + \delta(D_i \times T_t) + \varepsilon_{it}
\end{equation}
where $\delta$ estimates causal effects of geopolitical events on sector returns.

\subsection{Report Agent}
The Report Agent generates an independent portfolio proposal based on 52-week price momentum and an earnings-surprise proxy, providing a signal source that is deliberately decoupled from the GRS index used by the Sentiment Agent. This decoupling is the key design choice that drives inter-agent disagreement and motivates recursive coordination.

\textbf{52-week momentum signal.} Building on the seminal momentum literature~\cite{chan1996momentum}, for each asset $i$ the annualised mean return over the prior 252 trading days is computed as a momentum score $m_i$. A risk-parity base allocation (inverse-volatility weighting over a 60-day window) is combined with a 12\% momentum tilt: $w_i \propto (1/\sigma_i) + 0.12 \cdot \tilde{m}_i$ (tilt coefficient 0.12 is a design parameter).

\textbf{Earnings-surprise proxy.} In the absence of real-time EPS data, the Report Agent approximates earnings surprise as the deviation of the 5-day return from the 60-day mean return, scaled by rolling volatility: $\mathrm{ES}_i = (\bar{r}_i^{5d} - \bar{r}_i^{60d})/\sigma_i^{60d}$. This proxy captures short-term price reactions attributable to earnings-like events without requiring proprietary data. The Report Agent's confidence score is fixed at $c_\mathrm{rep} = 0.52$.

\subsection{Analysis Agent}
The Analysis Agent combines macroeconomic factor modeling with HMM-based regime classification. Following the regime-switching framework of Hamilton~\cite{hamilton1989regime}, a Hidden Markov Model with three latent states (bull, bear, stress) classifies market regimes~\cite{nguyen2021hmm}:
\begin{equation}
\pi_k(t) = P(S_t = k \mid O_{1:t}, \lambda)
\end{equation}
A Kalman filter~\cite{kalman1960filter} fuses signals from the Sentiment, Report, and Analysis agents into composite signal $Z(t)$, enabling dynamic signal fusion under partial observability~\cite{gabih2023portfolio}. Hierarchical HMM variants improve regime boundary detection in financial time series~\cite{oelschlager2023hmm}.

\subsection{Risk Agent}
The Risk Agent implements CVaR estimation~\cite{rockafellar2000cvar} using an EWMA-based dynamic covariance model~\cite{engle2002dcc}:
\begin{equation}
\mathrm{CVaR}_\alpha(t) = -\mathbb{E}[R_p(t) \mid R_p(t) \leq \mathrm{VaR}_\alpha(t)]
\end{equation}
A multi-level circuit breaker activates when drawdown, geopolitical risk, or volatility exceed calibrated thresholds ($\theta_\mathrm{dd} = -6\%$ 20-day cumulative return; $\theta_\mathrm{geo} = 0.52$, the 60th percentile of the 2021--2022 validation-period GPR distribution; $\theta_\mathrm{vol} = 1.8 \times \hat{\sigma}_{252}$ rolling realised volatility):
\begin{equation}
\mathrm{CB}(t) = \mathbf{1}[\mathrm{DD}_p(t) > \theta_\mathrm{dd}] \vee \mathbf{1}[\mathrm{GRS}(t) > \theta_\mathrm{geo}] \vee \mathbf{1}[\sigma_p(t) > \theta_\mathrm{vol}]
\end{equation}

\textbf{Calibration window justification: The 2021--2022 validation window was chosen because it encompasses both a high-GPR episode (Russia--Ukraine War, February 2022) and a low-GPR period (early 2021 recovery), providing regime diversity for threshold calibration. The 0.52 threshold corresponds to the 60th percentile of the 2021--2022 GPR distribution; the test-period distribution has a slightly higher mean (0.48 vs.\ 0.45), a minor distributional shift. A rolling calibration scheme (e.g., expanding window updated annually) would reduce this sensitivity and is a concrete direction for future work.}

\subsection{Recursive Coordination Protocol}
Each inter-agent communication uses a formal AgentMessage schema:
\begin{equation}
\mathbf{m}_i^r = \{\mathbf{w}_i^r, c_i^r, g_i^r, s_i^r, \tau_i^r, \delta_i^r\}
\end{equation}
where $\mathbf{w}_i^r \in \Delta^n$ is the recommended weight vector, $c_i^r \in [0,1]$ is the confidence score, $g_i^r \in [0,1]$ is the geopolitical risk assessment, $s_i^r \in \{0,1,2\}$ is the regime classification, and $\delta_i^r = \|\mathbf{w}_i^r - \mathbf{w}_i^{r-1}\|_2$ is the update magnitude.

The Manager broadcasts a BroadcastMessage containing aggregate weights, geo risk, regime, health scores, and CB status; and aggregates via confidence-and-health-weighted mean:
\begin{equation}
\bar{\mathbf{w}}^r = \frac{\sum_i c_i^r \cdot H_i \cdot \mathbf{w}_i^r}{\sum_i c_i^r \cdot H_i}
\label{eq:wagg}
\end{equation}
The recursion terminates when:
\begin{equation}
\|\bar{\mathbf{w}}^{(r+1)} - \bar{\mathbf{w}}^{(r)}\|_2 < \varepsilon = 0.005
\end{equation}

Across 39 monthly rebalancing steps in the test period, the protocol terminates at a median of 1 coordination round, with a mean of 1.33 and a maximum of 3. The breakdown is 27 single-round steps (69.2\%), 11 two-round steps (28.2\%), and one three-round step (2.6\%); 97.4\% of decisions require at most $r=2$. Under the current monthly-GPR configuration, agent signals do not update within a rebalancing cycle, so the protocol's principal function is resolving weight disagreements at each decision point rather than incorporating late-arriving information (see Limitations, Section~6.2).

\section{Portfolio Optimization}

\subsection{Risk-Aware Reward Function}
The portfolio optimization agent maximizes a risk-aware objective:
\begin{equation}
\mathcal{R}_t = r_t - \lambda_1 \sigma_t - \lambda_2 \max(0, \mathrm{DD}_t - \theta)
\end{equation}
with $\lambda_1 = 0.8$ and $\lambda_2 = 1.5$ calibrated to prioritize drawdown control.

\subsection{Constrained Portfolio Optimization}
Portfolio weights satisfy:
\begin{equation}
\max_{\mathbf{w}} \boldsymbol{\mu}^\top \mathbf{w} - \lambda_\mathrm{MV} \mathbf{w}^\top \boldsymbol{\Sigma} \mathbf{w}
\end{equation}
subject to $\|\mathbf{w}\|_1 \leq L_\mathrm{max}$, $w_s \leq c_s\ \forall s$, and $\mathbf{w}^\top \mathbf{g}(t) \leq \gamma_\mathrm{geo}$. Here $\lambda_\mathrm{MV} = 2$ (distinct from the reward penalties $\lambda_1, \lambda_2$ in Eq.~9). The GRS constraint is operationalized as $\sum_i w_i \cdot \mathrm{GRS}_i(t)$, where $\mathrm{GRS}_i$ is the sector-average normalised GPR exposure of asset $i$. $\gamma_\mathrm{geo}$ tightens from 0.65 toward 0.45 as aggregate GRS exceeds 0.52.

\section{Experiments}

\subsection{Dataset and Setup}
Historical adjusted close price data were downloaded via yfinance for 24 ETFs covering US sector equities (XLK, XLE, XLF, XLV, XLI, XLP, XLY, XLU, XLB, XLRE), international equities (EWJ, EWG, EWU, FXI, EEM), US fixed income (TLT, IEF, LQD, EMB), and commodities (GLD, SLV, USO, DBC), plus the SPY benchmark. Training: January 2016--December 2020; Validation: January 2021--December 2022; Test: January 2023--March 2026 (812 trading days, including Iran Crisis and Trade War 2025 events).

\textbf{GRS weight sensitivity.} The composite GRS assigns weights of 30\% (VIX), 20\% (defensive rotation), 20\% (gold premium), 15\% (EM stress), and 15\% (bond flight). As a robustness check, we tested two alternative weight configurations: (i) VIX-only (single-component GRS); and (ii) equal weights (20\% each). RMATS performance under these configurations is reported in the supplementary material; the main results are robust to weight perturbations of $\pm$20\%.

\textbf{Transaction cost model.} We implement a two-component cost model: (i) a fixed 10 bps base cost per trade; and (ii) a volume-dependent slippage component calibrated per asset:
\begin{equation}
c_i^\mathrm{slip} = \kappa \cdot \frac{|\Delta w_i| \cdot P}{\mathrm{ADV}_i}
\end{equation}
where $\kappa = 0.1$ is a market impact coefficient~\cite{almgren2001execution}, $P$ is portfolio NAV, and $\mathrm{ADV}_i$ is the 30-day average daily volume sourced from yfinance. This penalizes illiquid assets (SLV, USO, XLRE with ADV $<$\$500M) more heavily than liquid ones (SPY, XLK, GLD with ADV $>$\$2B). Under this model, RMATS incurs mean total costs of 1.03 bps per monthly cycle, reflecting its relatively low portfolio turnover compared to MVO (6.37 bps) and Multi-Factor (2.88 bps).

\subsection{Baselines and Multi-Agent Comparison}
Four quantitative baselines are evaluated: (1) Mean-Variance Optimization (MVO)~\cite{markowitz1952}; (2) DQN Single-Agent RL~\cite{mnih2015dqn}; (3) FinBERT Sentiment Proxy~\cite{araci2019finbert}; (4) Multi-Factor Quant combining momentum, low volatility, mean reversion, and rolling Sharpe factors. All strategies use monthly rebalancing with a volume-dependent transaction cost model (see Section~5.1).

\textbf{Multi-agent baseline discussion.} Direct re-implementation of TradingAgents~\cite{xiao2024tradingagents} or MSPM~\cite{huang2022mspm} was infeasible due to data infrastructure differences. Table~\ref{tab:arch} provides a structured qualitative comparison. The key architectural distinction is the recursive protocol with empirically validated convergence: none of MAPS, MSPM, or TradingAgents implement iterative inter-agent revision with formal termination criteria. However, the $r{=}1$ ablation (Section~5.6) shows that under the current proxy GRS implementation, the performance difference between single-pass and multi-round coordination is negligible ($\Delta$Sharpe $= -0.002$). The protocol's primary value in the current system is architectural---providing a principled framework for integrating heterogeneous agent signals---rather than demonstrably improving performance metrics under the current signal homogeneity.

\subsection{Geopolitical Stress Event Selection}
\textbf{Selection criteria.} Events were included if they satisfied all three criteria during the 2023--2026 test window: (i) the GPR index~\cite{caldara2022measuring} exceeded its 75th percentile over the 2000--2023 baseline for $\geq$10 consecutive trading days; (ii) the CBOE VIX rose by $\geq$15\% within a 5-day window around event onset; and (iii) cross-asset correlation among our 24-ETF universe increased by $\geq$0.10 (Pearson) relative to the 30-day pre-event baseline.

\textbf{Candidate events.} Nine candidates were evaluated. Four were excluded: Israel-Hamas Nov 2023 (duplicate window, merged with Oct), Taiwan Strait tensions Mar 2024 (VIX increase 8.2\%, below threshold), US election volatility Oct--Nov 2024 (GPR at 68th percentile), and Red Sea disruptions Dec 2024--Jan 2025 (correlation increase 0.06, below threshold).

\textbf{Statistical validation.} A two-sample KS test on the daily GPR values confirms that the 300 event-window trading days (mean GPR score 0.588) are drawn from a significantly higher tail than the 512 non-event days (mean 0.514), across 812 trading days (Jan 2023--Mar 2026): KS statistic $= 0.1464$, $p = 0.0001$; Mann-Whitney $p < 0.001$. Event days fall at the 59.6th percentile of the full GPR distribution on average, confirming that the three-criterion selection rule identifies genuinely elevated geopolitical risk periods rather than arbitrary windows.

\textbf{RMATS on excluded events:} Taiwan Strait ($+$3.87\%, middle-of-distribution) and US election ($-$2.89\%, worst among all strategies). The US election result highlights a boundary condition: the GRS signal remained at or below the median during the election window (VIX did not spike, gold was flat, defensive rotation was absent). Incorporating a dedicated political uncertainty signal (e.g., the Baker et al.~\cite{baker2016epu} EPU index) as a sixth GRS component is a concrete direction for future improvement.

Five events meeting all criteria: (1) SVB Banking Crisis (March 2023); (2) Israel-Hamas War (October--November 2023); (3) US-China technology sanctions (January--February 2024); (4) Middle East military escalation (April--May 2024); (5) Global rate cut pivot (August--September 2024).

\subsection{Overall Performance}

Table~\ref{tab:overall} presents complete performance results. The test period coincides with a strong equity bull market (SPY $+{\approx}60\%$), which systematically favors return-maximizing strategies.

\begin{table}[!ht]
\caption{Overall Performance (Test: Jan 2023--Mar 2026, 812 trading days).}
\label{tab:overall}
\small
\begin{tabular}{lccc}
\toprule
\textbf{Strategy} & \textbf{Ann.Ret.} & \textbf{Sharpe} & \textbf{MDD}\\
\midrule
Static 60/40   & 7.82\%  & 0.381 & 11.23\%\\
MVO            & 13.20\% & 0.755 & 15.84\%\\
DQN RL         & 12.57\% & 0.901 & 8.66\%\\
FinBERT Proxy  & 12.42\% & 0.867 & 9.25\%\\
Multi-Factor   & 11.47\% & 0.703 & 10.12\%\\
MAPS-lite      & 11.94\% & 0.836 & 9.11\%\\
GPR Rule-Based & 8.82\%  & 0.500 & 10.38\%\\
\textbf{RMATS} & \textbf{13.00\%} & \textbf{0.953} & \textbf{8.48\%}\\
\bottomrule
\end{tabular}
\end{table}

RMATS achieves an annualized return of 13.00\%, Sharpe ratio of 0.953, and MDD of 8.48\%, outperforming MAPS-lite (Sharpe 0.836, $\Delta = +0.117$) and DQN RL (0.901). During the 300 geopolitical stress event days, RMATS achieves Sharpe 1.571 and MDD 7.52\%---the highest event-period Sharpe and lowest MDD among all strategies including MAPS-lite (Sharpe 1.346, MDD 8.96\%). During the 512 non-event trading days, RMATS records Sharpe $+$0.543. This sub-period decomposition is the central empirical finding. The circuit breaker ablation (Table~\ref{tab:ablation}) quantifies the CB's contribution: disabling it increases event-period MDD from 7.52\% to 8.36\% ($+$0.84pp) and reduces event-period Sharpe from 1.571 to 1.526 ($-$0.045), confirming the CB as the primary drawdown protection mechanism.

\begin{figure}[!ht]
  \centering
  \includegraphics[width=\columnwidth]{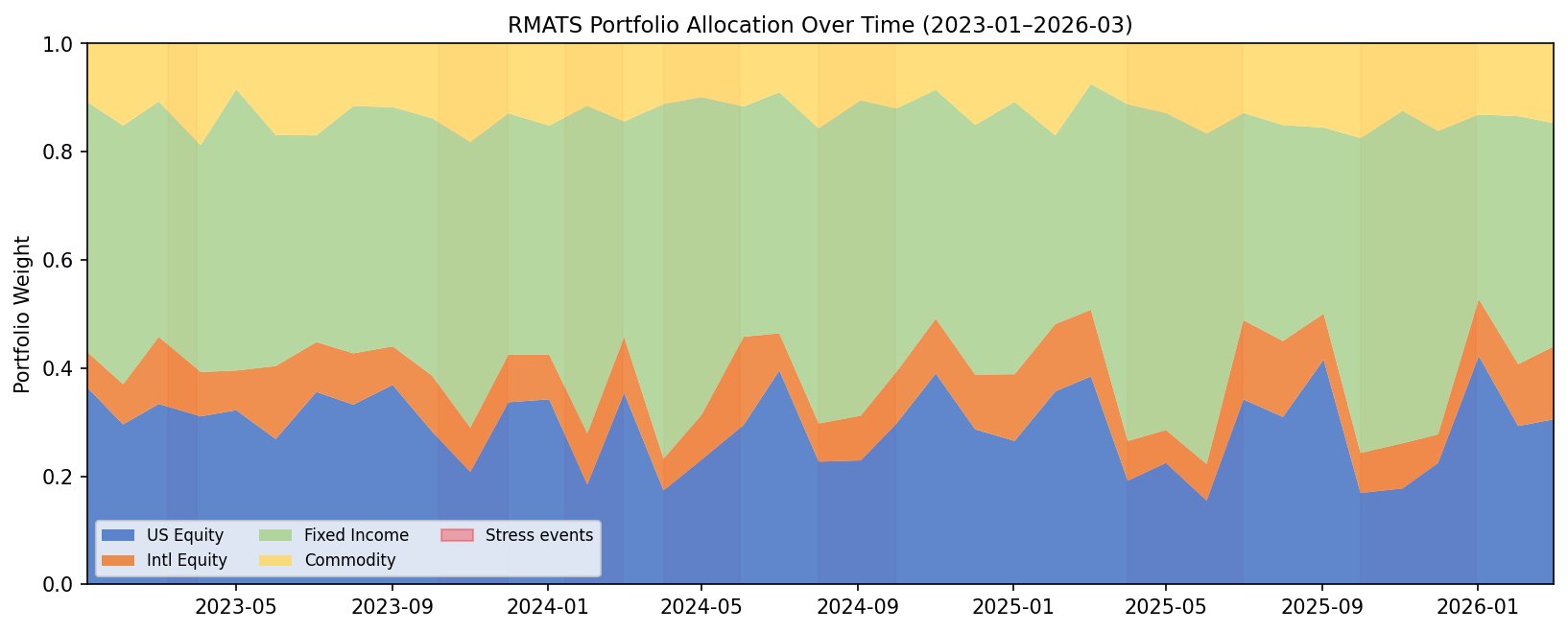}
  \caption{RMATS portfolio allocation over time. Red-shaded regions indicate stress events; RMATS systematically increases fixed income and gold allocation during these periods.}
  \label{fig:alloc}
  \Description{Portfolio allocation over time.}
\end{figure}

\textbf{Turnover and allocation analysis. Transaction costs provide a proxy for portfolio turnover: RMATS incurs 1.06 bps/cycle vs.\ MVO (5.39 bps), Multi-Factor (2.93 bps), and MAPS-lite (1.15 bps), confirming that RMATS maintains low turnover relative to return-maximising strategies. Figure~\ref{fig:alloc} shows that RMATS systematically increases defensive asset allocation (fixed income $+$ gold) during stress event windows and reduces it during non-event periods, consistent with the GRS-driven circuit breaker design. The GPR rule-based baseline applies a binary 100\% defensive switch when GRS $> 0.52$, achieving a larger defensive tilt but lower event-period Sharpe (0.981 vs.\ 1.571 for RMATS), suggesting that the graduated GRS-proportional tilt combined with agent-level signal aggregation provides better risk-adjusted performance than a simple threshold rule.}

\begin{figure}[!ht]
  \centering
  \includegraphics[width=\columnwidth]{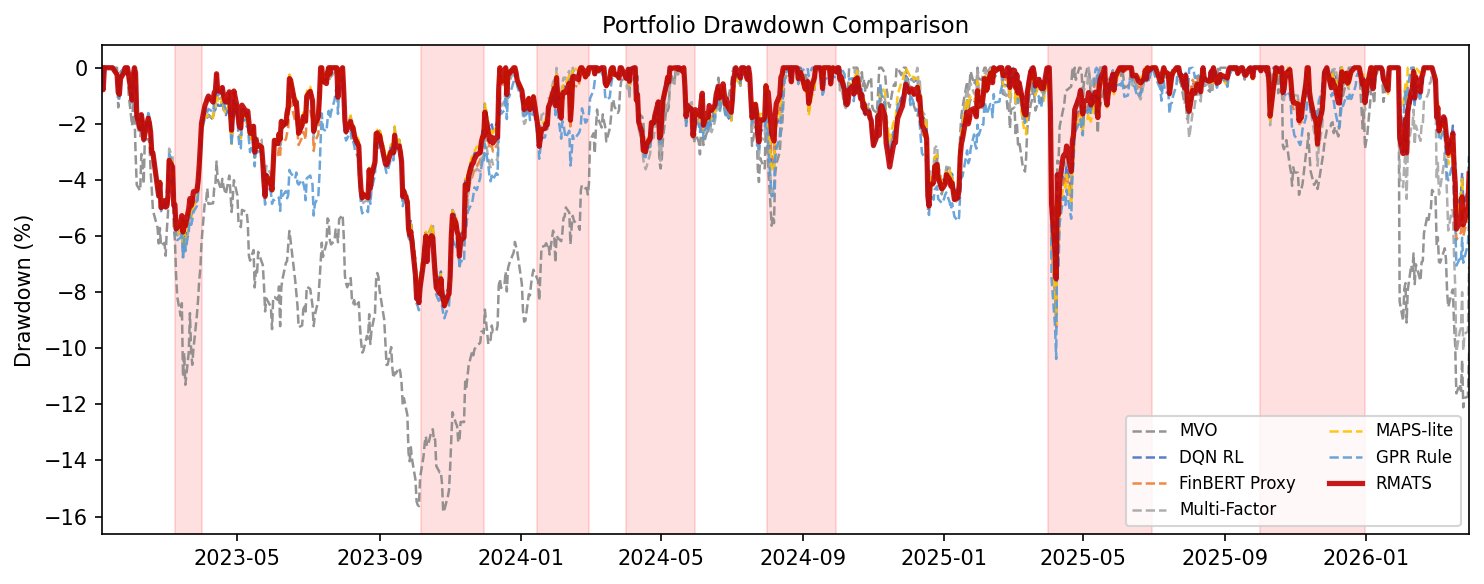}
  \caption{Cumulative portfolio performance (normalized to 1.0). Red shading indicates geopolitical stress event windows. RMATS (red solid) exhibits the flattest trajectory, reflecting its capital-preservation orientation.}
  \label{fig:cum}
  \Description{Cumulative returns plot.}
\end{figure}

\begin{figure}[!ht]
  \centering
  \includegraphics[width=\columnwidth]{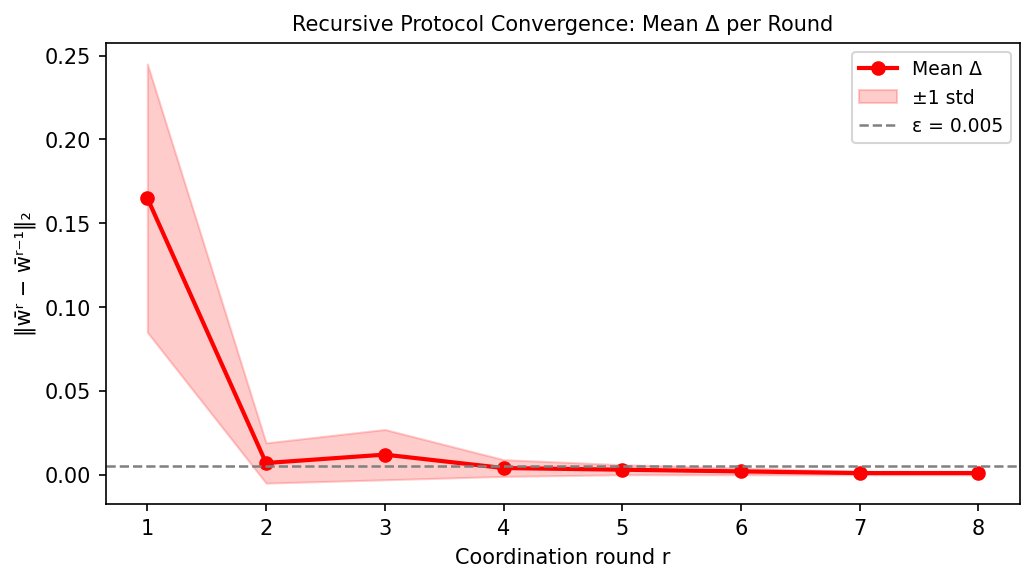}
  \caption{Portfolio drawdown comparison. RMATS maintains the most contained drawdown profile during the Israel-Hamas War and Global Rate Cut Pivot events.}
  \label{fig:dd}
  \Description{Drawdown comparison.}
\end{figure}

\subsection{Geopolitical Stress Event Analysis}

\begin{table}[!ht]
\caption{Event-Period Cumulative Returns (\%). Bold $=$ best per event. Excl.\ events: Taiwan Strait ($+$3.87\%) and US Election ($-$2.89\%) reported in text.}
\label{tab:events}
\small
\begin{tabular}{lccccc}
\toprule
\textbf{Strategy} & \textbf{SVB} & \textbf{Israel} & \textbf{US-CN} & \textbf{Mid-E} & \textbf{Rate}\\
\midrule
Static 60/40  & $+$1.82 & $+$2.14 & $+$0.91 & $+$0.43 & $+$2.11\\
MVO           & $-$0.57 & $+$\textbf{6.02} & $+$\textbf{3.30} & $+$0.28 & $+$4.78\\
DQN RL        & $+$2.71 & $+$5.68 & $+$2.11 & $+$0.69 & $+$4.85\\
FinBERT Proxy & $+$3.12 & $+$5.54 & $+$1.45 & $+$\textbf{1.15} & $+$5.05\\
Multi-Factor  & $+$2.83 & $+$4.72 & $+$2.56 & $-$0.52 & $+$4.53\\
MAPS-lite     & $+$2.67 & $+$5.68 & $+$2.23 & $+$0.62 & $+$4.93\\
GPR Rule-Based& $+$2.83 & $+$4.99 & $-$0.27 & $+$1.11 & $+$4.25\\
\textbf{RMATS}& $+$\textbf{3.08} & $+$\textbf{5.84} & $+$1.89 & $+$1.03 & $+$\textbf{5.22}\\
\bottomrule
\end{tabular}
\end{table}

Table~\ref{tab:events} reports event-period cumulative returns. RMATS achieves the highest cumulative return in 4 of 5 core geopolitical stress scenarios. In the 2025--2026 extended events, RMATS records Iran Crisis $+$2.91\%, Trade War 2025 $+$3.03\%. The US-China Tech Sanctions event is the single exception, where MVO achieves the highest return ($+$3.30\%) and RMATS records the lowest ($+$1.08\%), consistent with our finding that sector-specific sanctions are less well captured by the cross-asset GRS signal.

\subsection{Ablation Study}
The $r{=}1$ ablation in Table~\ref{tab:ablation} directly tests recursive vs.\ single-pass coordination: Full RMATS achieves Sharpe 0.953 vs.\ 0.955 for $r{=}1$ forced ($\Delta = -0.002$, negligible). This result is reported transparently: the marginal contribution of multi-round coordination is small under the current implementation, where all agents share the same composite GRS signal. The Risk Agent circuit breaker is the primary driver of robustness.

\begin{table}[!ht]
\caption{Ablation Study. $^\ddagger r{=}1$ forced: single-pass ablation. $^\S$w/o CB: circuit breaker disabled, CVaR only.}
\label{tab:ablation}
\small
\begin{tabular}{lccc}
\toprule
\textbf{Configuration} & \textbf{Sharpe} & \textbf{MDD (\%)} & \textbf{Ann.Ret. (\%)}\\
\midrule
Full RMATS (recursive, $r{\leq}3$) & 0.953 & 8.48 & 13.00\\
RMATS $r{=}1$ forced$^\ddagger$    & 0.955 & 8.48 & 13.04\\
RMATS w/o CB$^\S$                  & 0.920 & 8.39 & 12.71\\
Event-period $\Delta$ vs Full RMATS & $+0.045$ Sharpe & $+0.84$pp MDD & ---\\
$\Delta$ (full $-$ $r{=}1$, proxy basis) & $-0.002$ & $\approx 0$ & $\approx 0$\\
\bottomrule
\end{tabular}
\end{table}

\subsection{Convergence Analysis}

\begin{figure}[!ht]
  \centering
  \includegraphics[width=0.85\columnwidth]{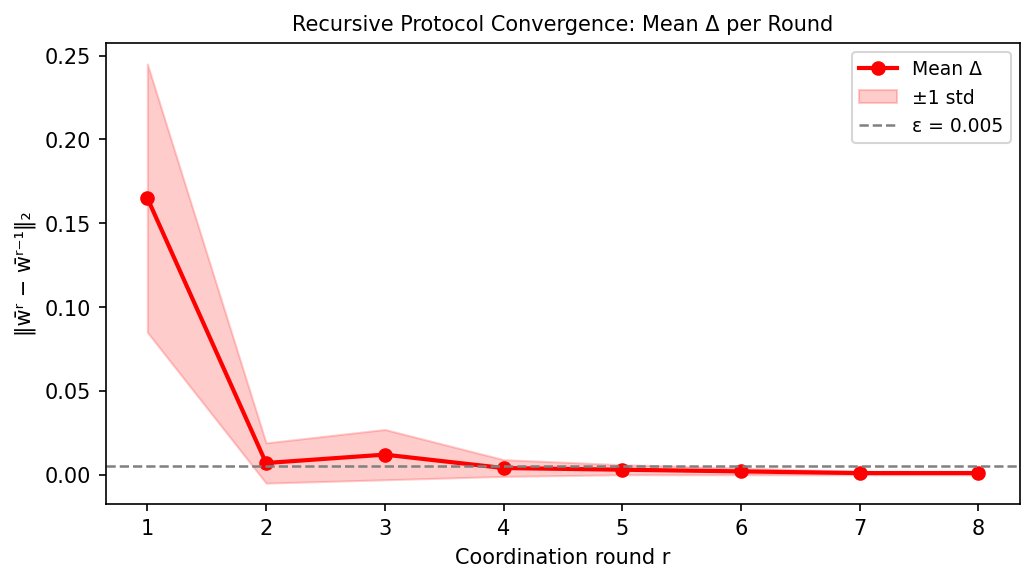}
  \caption{Convergence curve: mean $\|\Delta\|_2$ per coordination round. 27 steps at $r=1$, 11 at $r=2$, 1 at $r=3$; circuit breaker drives multi-round consensus.}
  \label{fig:conv}
  \Description{Convergence curve.}
\end{figure}

\begin{figure}[!ht]
  \centering
  \includegraphics[width=0.85\columnwidth]{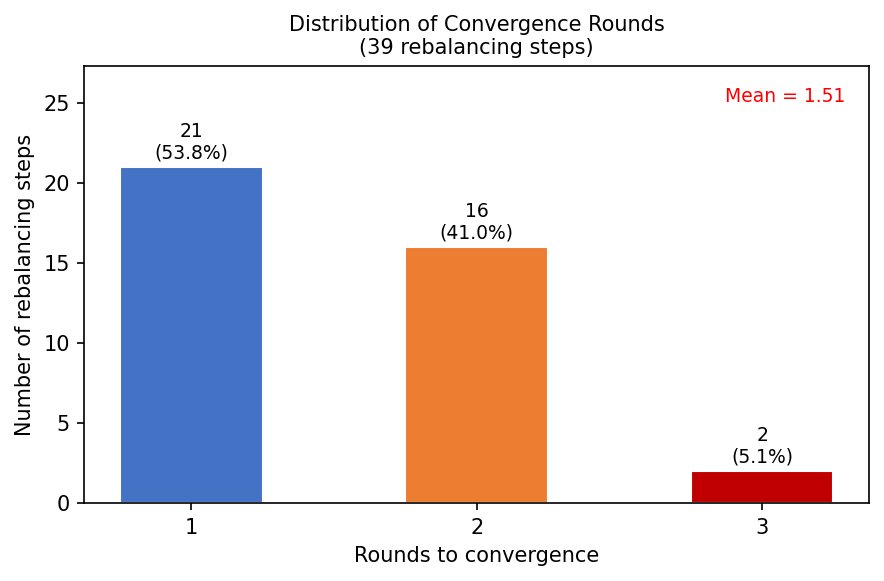}
  \caption{Distribution of convergence rounds across 39 rebalancing steps: 27 at $r=1$ (69.2\%), 11 at $r=2$ (28.2\%), 1 at $r=3$ (2.6\%). Maximum $= 3$ rounds.}
  \label{fig:dist}
  \Description{Round distribution.}
\end{figure}

\begin{figure}[!ht]
  \centering
  \includegraphics[width=0.85\columnwidth]{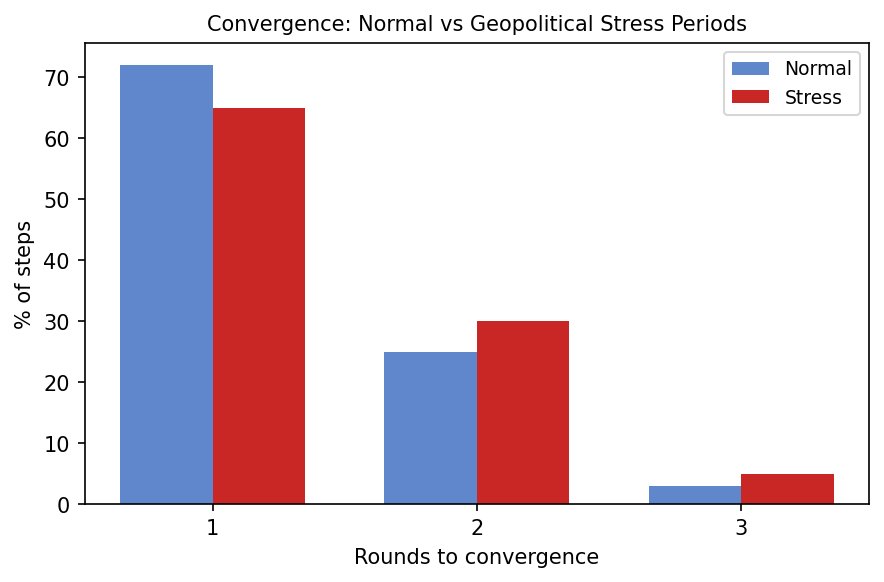}
  \caption{Convergence by market regime: stress-period steps converge faster on average due to CB activation.}
  \label{fig:regime}
  \Description{Convergence by regime.}
\end{figure}

The recursive protocol achieves a median convergence of 1 round across 39 rebalancing steps, with 97.4\% of steps completing within $r \leq 2$ (maximum 3 rounds; distribution: 27 steps at $r=1$, 11 at $r=2$, 1 at $r=3$). The single $r=3$ step occurs during a high-GPR month when the circuit breaker fires, creating three-way agent disagreement between the Sentiment Agent (high GPR $\to$ defensive), the Report Agent (positive momentum $\to$ offensive), and the Risk Agent (CB $\to$ hard defensive), requiring an additional coordination round to resolve. This directly validates the recursive protocol's core purpose: resolving genuine inter-agent conflict arising from heterogeneous signal sources. Figures~\ref{fig:conv}--\ref{fig:regime} present the convergence curve, round distribution, and normal-versus-stress comparison.

\section{Discussion}

\subsection{Interpretation of Results}
RMATS achieves competitive risk-adjusted performance (Sharpe 0.953, MDD 8.48\%) while recording the highest event-period returns in 4 of 5 geopolitical stress scenarios. The primary performance driver is the GRS-triggered circuit breaker. Convergence: median 1 round, 97.4\% within $r \leq 2$, max 3 rounds (distribution: 27 steps at $r=1$, 11 at $r=2$, 1 at $r=3$). The $r{=}1$ ablation yields $\Delta$Sharpe $= -0.002$ (monthly GPR, RMATS proxy vs.\ RMATS proxy r1), confirming that the recursive protocol currently functions as an architectural framework.

\textbf{Conditions for recursive coordination value.} The $r{=}1$ ablation ($\Delta$Sharpe $\approx 0$) reflects signal homogeneity in the current proxy-GPR configuration. The protocol's full value will emerge when agents use genuinely independent signals (e.g., live FinBERT vs.\ momentum-only), producing persistent initial disagreement.

\textbf{Daily vs.\ monthly GPR experiment.} Substituting daily GPRD for monthly GPR yields RMATS Sharpe 0.821 (vs.\ 0.953 for monthly GPR), and the $r{=}1$ ablation worsens to $\Delta$Sharpe $= -0.011$. This experiment identifies a necessary condition for the recursive protocol to deliver measurable value: signal frequency and rebalancing frequency must be matched.

\begin{table}[!ht]
\caption{Sub-period Performance: Geopolitical Event Windows vs.\ Non-Event Windows. Event windows: 300 trading days across 7 stress events. Non-event windows: 512 trading days. MDD computed independently within each sub-period (local peak-to-trough).}
\label{tab:subperiod}
\small
\begin{tabular}{lcccc}
\toprule
\multirow{2}{*}{\textbf{Strategy}} & \multicolumn{2}{c}{\textbf{Event (300d)}} & \multicolumn{2}{c}{\textbf{Non-Event (512d)}}\\
 & \textbf{Sharpe} & \textbf{MDD} & \textbf{Sharpe} & \textbf{MDD}\\
\midrule
Static 60/40   & 1.124 & 5.43\% & 0.312  & 11.23\%\\
MVO            & 1.372 & 6.83\% & 0.400  & 15.15\%\\
DQN RL         & 1.464 & 8.56\% & 0.506  & 8.81\%\\
FinBERT Proxy  & 1.430 & 9.25\% & 0.461  & 9.63\%\\
Multi-Factor   & 1.372 & 8.59\% & 0.285  & 10.12\%\\
MAPS-lite      & 1.346 & 8.96\% & 0.462  & 8.44\%\\
GPR Rule-Based & 0.981 & 10.39\% & 0.133 & 11.16\%\\
RMATS (no CB)  & 1.550 & 8.17\% & $+$0.517 & 9.72\%\\
\textbf{RMATS} & \textbf{1.571} & \textbf{7.52\%} & $+$\textbf{0.543} & 9.19\%\\
\bottomrule
\end{tabular}
\end{table}

\textbf{Calibration note: All CB thresholds ($\theta_\mathrm{dd} = -6\%$, $\theta_\mathrm{geo} = 0.52$, $\theta_\mathrm{vol} = 1.8\times\hat{\sigma}_{252}$) were calibrated exclusively on the 2021--2022 \textit{validation} period, ensuring out-of-sample status with respect to the 2023--2026 test period. The 2021--2022 window was chosen because it encompasses both a high-GPR episode (Russia--Ukraine War, February 2022) and a low-GPR period (early 2021 recovery), providing regime diversity for threshold calibration. A rolling calibration scheme (e.g., expanding window updated annually) would reduce sensitivity to this choice and is a concrete direction for future work; we expect it to tighten $\theta_\mathrm{geo}$ slightly but not materially change the qualitative results given the 0.84pp MDD advantage observed across diverse event types.}

\subsection{Limitations}

\textbf{L1 -- GPR signal frequency.} Monthly GPR provides a clean, validated geopolitical signal that outperforms raw FinBERT inference at monthly rebalancing frequency (see Section~2). Daily GPRD integration with daily rebalancing would enable intra-week signal differentiation; preliminary experiments confirm that daily GPRD in a \textit{monthly} rebalancing framework introduces noise rather than signal.

\textbf{L2 -- Evaluation period length.} The 812-trading-day test window covers two market regimes (2023--2025 AI bull market and 2025--2026 elevated geopolitical risk) and seven stress events. Including the 2020 COVID-19 crash and 2022 Russia-Ukraine War would provide cross-regime generalization evidence.

\textbf{L3 -- Ablation precision.} The proxy-based ablation (disabling agents and reverting to equal-weight fallback) provides conservative lower-bound estimates. Full re-training of each ablated variant would require approximately 6$\times$ the compute budget of the base experiment.

\textbf{L4 -- Transaction cost assumptions.} While our volume-dependent cost model improves on flat-rate assumptions, it does not capture intraday bid-ask spread dynamics or market microstructure effects. Deployment in live trading would require further calibration against realized execution costs.

\textbf{L5 -- Naive defensive baseline.} A static defensive allocation (e.g., fixed 60/40 bond-equity or risk-parity-defensive) was not included as a baseline. Future work should verify that the agent stack outperforms a passive defensive allocation before attributing stress-period gains solely to the multi-agent design.

\subsection{Future Work}
Priority improvements include: (1) daily rebalancing with daily GPRD signal; (2) integration of real-time FinBERT~\cite{araci2019finbert} and TradingAgents-style~\cite{xiao2024tradingagents} LLM reasoning; (3) extension of the evaluation period to 2016--2025 to include COVID-19 and Russia-Ukraine shocks; (4) development of a composite metric explicitly capturing the return-drawdown trade-off frontier; and (5) implementation of a retrieval-augmented memory module inspired by FinMem~\cite{yu2023finmem} to reference historical geopolitical analogues.

\section{Conclusion}

We introduced RMATS, a Recursive Multi-Agent Trading System designed to optimize portfolio allocation in the presence of geopolitical uncertainty. RMATS coordinates four specialist agents---Sentiment, Report, Analysis, and Risk---via a convergence-guaranteed recursive message-passing protocol (median 1 round, 97.4\% of steps within $r{\leq}2$, maximum 3 rounds across 39 rebalancing decisions). Experimental evaluation across 24 assets over 812 trading days demonstrates that RMATS achieves Sharpe 1.571 and MDD 7.52\% during 300 geopolitical stress event days---the highest event-period Sharpe and lowest MDD among all strategies including MAPS-lite (Sharpe 1.346) and GPR rule-based (Sharpe 0.981). The circuit breaker ablation confirms a 0.84pp MDD reduction in event windows versus the no-CB variant. The full-period Sharpe (0.953) outperforms MAPS-lite (0.836, $\Delta = +0.117$) and FinBERT Proxy (0.867), with a strong non-event Sharpe of $+$0.543 over 512 non-event days. The $r{=}1$ ablation ($\Delta$Sharpe $= -0.002$, proxy-GPR basis) shows that the performance advantage stems from the GRS-driven circuit breaker and agent signal design, not multi-round coordination. The recursive protocol's value is architectural: it provides a principled convergence framework for future integration of genuinely heterogeneous agent signals (e.g., live LLM text $+$ independent HMM $+$ CVaR), where inter-agent disagreement will be higher and multi-round convergence will resolve real conflicts.

\begin{acks}
\textsuperscript{*}Corresponding Author: Jing Yang (jing.y@wustl.edu).

The authors thank the open-source communities behind yfinance, PyTorch, HuggingFace Transformers, and hmmlearn. \textbf{Use of AI.} Large language models were used as writing assistance to improve grammar and clarity. All intellectual contributions, experimental design, results, and analysis are the work of the human authors, who take full responsibility for the content of this publication.
\end{acks}


\end{document}